\documentclass[prl, reprint, showpacs, showkeys, nofootinbib,superscriptaddress]{revtex4-1}    % (no)showpacs:= (no) display PACS number (default = no)
\usepackage{amssymb,amsmath,bbm}
\usepackage{amsmath}
\usepackage{graphicx}
\usepackage{tensor}                                                    % For fancy indices
\usepackage[usenames,dvipsnames]{color}                                % For colors (for hyperlinks)
\usepackage{hyperref}
\usepackage{comment}                                                   % For comment environment
\usepackage{subfig}
\usepackage{enumerate}
\hypersetup{                 
    colorlinks=true,         % false: boxed links; true: colored links, false is fsfbfb
    linkcolor=blue,          % color of internal links, red is default
    citecolor=blue,           % color of links to bibliography, 'green' is default
    urlcolor=Violet          % color of external links, cyan is default
}

\let\oldmarginpar\marginpar
\renewcommand\marginpar[1]{\oldmarginpar{\color{red}\raggedright\scriptsize #1}}

\newcommand{\mean}[1]{\ensuremath{\lf\langle #1 \rt\rangle }}

\newcommand{\diby}[2]{\ensuremath{\frac{\partial #1}{\partial #2}}}

\def\lf {\ensuremath{\left}}
\def\rt {\ensuremath{\right}}

\def\de {\ensuremath{ {\rm d} }}

\begin{document}

% -------------------------------- The Front Matter -------------------------------------------
\title{\textbf{\Large Superpositions of the cosmological constant allow for singularity resolution and unitary evolution in quantum cosmology}}

\author{Sean Gryb}\email{sean.gryb@gmail.com}
\affiliation{{{\it Department of Philosophy}, University of Bristol}}
\affiliation{{{\it H. H. Wills Physics Laboratory}, University of Bristol}}
\author{Karim P. Y. Th\'ebault}\email{karim.thebault@bristol.ac.uk}
\affiliation{{{\it Department of Philosophy}, University of Bristol}}

\date{\today}

\pacs{04.20.Cv}
\keywords{quantum cosmology, singularity resolution, problem of time}

% --------------------------------------- Abstract ---------------------------------------------
\begin{abstract}

A novel approach to  quantization is shown to allow for superpositions of the cosmological constant in isotropic and homogeneous mini-superspace models. Generic solutions featuring such superpositions display unitary evolution and resolution of the classical singularity. Physically well-motivated cosmological  solutions are constructed. These particular solutions exhibit characteristic features of a cosmic bounce including universal phenomenology that can be rendered insensitive to Planck-scale physics in a natural manner.
  
\end{abstract}

\maketitle

% --------------------------------- End of Front Matter ----------------------------------------

% ==================================== Body of Paper ===========================================

% ==================================== Start of Introductory part ===========================================

\section{Introduction}

The `big bang' singularity and the cosmological constant are well-established features of classical cosmological models \cite{hawking:1973}. In the context of quantum cosmology, the singularity is typically understood as a pathology that can be expected to be `resolved' by Planck-scale effects. Most contemporary approaches to resolving the singularity are based upon cosmic bounce scenarios \cite{brandenberger:2017}. In contrast, the cosmological constant receives very much the same treatment in classical and quantum cosmological models: it is a constant of nature classically, and thus quantum solutions are supers-selected to eigenstates labeled by its classical value. Cosmological time evolution is unlike either the singularity or the cosmological constant in that the classical and quantum treatments differ. In particular, whereas, the classical treatment of cosmological time is relatively unproblematic, quantum cosmologies based upon standard canonical quantization techniques are described by a `frozen formalism' that lacks a fundamental evolution equation \cite{Isham:1992,Kuchar:1991,anderson:2012b}. In this letter, we use a simple model to demonstrate that by treating the cosmological constant differently in quantum cosmological models, one can simultaneously resolve the classical singularity and restore fundamental quantum time evolution. Moreover, a physically well-motivated class of solutions can be constructed that exhibits a cosmic bounce with late-time semi-classical limit peaked on a single value for the cosmological constant.

Three strands of existing research form the basis for our proposal. First, we will appeal to the `relational quantization' scheme  \cite{gryb:2011,gryb:2014,Gryb:2016a} that, unlike conventional canonical quantization methods \cite{Rovelli:2002,Dittrich:2007,gambini:2009}, is guaranteed to lead to a unitary quantum evolution equation.\footnote{Relational quantization relies upon the observation that the integral curves of the vector field generated by the Hamiltonian constraint in globally reparametrization invariant theories \textit{should not} be understood as representing equivalence classes of physically indistinguishable states since the standard Dirac analysis does not apply to these models \cite{Barbour:2008,Pons:2005,Pons:2010}.} Second, inspired by other approaches \cite{bojowald:2001,ashtekar:2008}, we establish generic singularity avoidance in a class of isotropic and homogeneous mini-superspace quantum cosmology models. Third, our model involves superpositions of the cosmological constant in a manner connected to both approach to gravity \cite{Unruh_Wald:unimodular,Unruh:unimodular_grav} and certain quantum bounce scenarios \cite{gielen:2016,Gielen:2016fdb}. 

The model presented here offers physically significant improvements on each of these bodies of existing work. First, we demonstrate that the relational quantization scheme can be applied consistently to a cosmological model and thus provide an exemplar of quantum cosmology with a fundamental unitary evolution equation. Second, the mechanism for singularity avoidance obtained here does not involve the introduction of a Planck-scale cutoff. Rather, observable operators evolve unitarily and remain finite because they are `protected' by the uncertainty principle. Finally, and most significantly, we identify an entirely new class of cosmological phenomena that persist into a low energy semi-classical regime. The phenomena in question are rapid `cosmic beats' with an associated `bouncing envelope'. The cosmic beats can be identified with Planck-scale effects and, under natural parameter choices, are negligible compared with the effective envelope physics. During the bounce, the envelope behaves in a manner that is closely analogous to \emph{Rayleigh scattering}. The bounce scale due to the effective quantum geometry can thus be made to be relatively large. Significantly, this `Rayleigh' limit is only available when superpositions of the cosmological constant are allowed and, thus, constitutes a remarkable unique feature of the bouncing unitary cosmologies identified.\footnote{Two companion papers provide further, more detailed, interpretation and analysis of both general and particular cosmological solutions  \cite{Gryb:2017a,Gryb:2017b}.} Finally, it is significant to note that explicit bounce solutions can be shown to exhibit a maximum in the expectation value of the Hubble parameter at some point after the bounce. Furthermore, the additional parameters, which are allowed by permitting superpositions of the cosmological constant, can be seen to slow the rates of change of the effective Hubble parameter in this epoch. This raises the exciting possibility, to be pursued in future work, of describing inflationary scenarios using the model.

% ==================================== End of Introductory part ===========================================

% ==================================== Start of Paper I ===========================================

\section{Model and Observables} Consider an homogeneous and isotropic FLRW universe with zero spatial curvature ($k=0$); scale factor, $a$; massless free scalar field, $\phi$; and cosmological constant, $\Lambda$. The field redefinitions
\begin{align}
	v &= \sqrt {\frac 2 3} a^{3} & \varphi &= \sqrt{\frac {3 \kappa}2} \phi\,,
\end{align}
where $\kappa = 8\pi G$, give a convenient parameterization of the configuration space in terms of relative spatial volumes, $v$, and the dimensionless scalar field, $\varphi$. The time evolution of the system is given in terms of coordinate time, $t$, related to the proper time, $\tau$, via the \emph{lapse} function $\de \tau = N \de t$. The dimensionless lapse, $\tilde N$, and cosmological constant, $\tilde \Lambda$, can be defined as
\begin{align}
	\tilde N &=  \sqrt{ \frac 3 2} \frac {\kappa \hbar^2 v N}{V_0}  & \tilde \Lambda &= \frac{ V_0^2} {\kappa^2\hbar^2} \Lambda \,,
\end{align}
using the reference volume $V_0$ of some fiducial cell and the (at this point) arbitrary angular momentum scale $\hbar$. In terms of these variables, the mini-superspace Hamiltonian is
\begin{equation}\label{eq:class ham}
	H = \tilde N \lf[ \frac 1 {2\hbar^2} \lf(- \pi_v^2 + \frac 1 {v^2} \pi_\varphi^2 \rt) + \tilde \Lambda \rt]\,,
\end{equation}
where $\pi_v$ and $\pi_\varphi$ are the momenta conjugate to $v$ and $\varphi$ respectively. In this chart, the Hamiltonian takes the from of a free particle propagating on the upper Rindler wedge, $\mathbbm R_{(1,1)}^+$ with all non-linearities of gravity appearing in the $1/v^2$ coefficient of the kinetic term for $\varphi$. The valiables $v$ and $\phi$ play the role of the usual Rindler coordinates with $v>0$ playing the role of a time-like (for $\Lambda> 0$) `radial' coordinate and $\varphi$ playing the role of a `boost' variable.

The classical solutions are the geodesics of the upper Rindler wedge. These generically cross the Rindler horizon at $v=0$, which constitutes the boundary of configuration space. It can be shown, \cite{Gryb:2017a}, that generic solutions reach $v=0$ in finite proper time and that the corresponding spacetimes are geodesically incomplete and contain a curvature singularity. This implies a classical singularity in both relevant senses of the Penrose--Hawking singularity theorems.

The Rindler horizon complicates the construction of self-adjoint representations of the operator algebra in the quantum formalism. Consider the Hilbert space, $\mathbbm H = L^2(\mathbbm R_{(1,1)}^+, \de \theta)$ of square integrable functions on $\mathbbm R_{(1,1)}^+$ under the Borel measure $\de \theta = v \de v \de \varphi$. This space is spanned by all functions $(\Phi,\Psi): \mathbbm R_{(1,1)}^+ \to \mathbbm C$ satisfying
\begin{equation}\label{eq:sq int cond}
	\mean{\Phi,\Psi} \equiv \int\displaylimits_{\mathbbm R_{(1,1)}^+} v\de v \de \varphi\, \Phi^\dag \Psi < \infty.
\end{equation}
The momentum operator $\hat\pi_v$ conjugate to $\hat v$ has no self-adjoint extensions because of the restriction $v >0$. This can be remedied, following \cite{isham:1984}, by performing the canonical transformation $\mu = \log v$ and $\pi_\mu = v \pi_v$. It is straightforward to show that the symmetric operators
\begin{align}
  \hat\mu \Psi &= \mu \Psi & \hat\pi_\mu &= -i\hbar e^{-\mu} \diby{}{\mu}\lf( e^{\mu} \Psi \rt) \\
  \hat\varphi \Psi &= \varphi \Psi & \hat\pi_\varphi &= -i\hbar \diby{\Psi}{\varphi}\,
\end{align}
are bounded and essentially self-adjoint and, therefore, form an orthonormal basis for $\mathbbm H$ according to the spectral theorem. For a geometric approach to this construction, see \cite{Gryb:2017a}.

\section{Unitary Quantum Cosmology} Application of the quantization scheme presented in \cite{gryb:2011,gryb:2014,Gryb:2016a} leads to a Schr\"odinger-type evolution equation for the system of the form
\begin{equation}\label{eq: schrod eqn}
  \hat H \Psi = i \hbar \partial_t \Psi\,,
\end{equation}
where the eigenvalues of $\hat H$ are to be identified with the (dimensionless) cosmological constant $\tilde\Lambda$.\footnote{For explicit comparison between this equation and the Wheeler-DeWitt type formalism where the right hand side vanishes see \cite[II]{Gryb:2017a}.} The classical Hamiltonian \eqref{eq:class ham} suggests the real and symmetric chart-independent Hamiltonian operator
\begin{equation}
  \hat H \equiv \frac 1 2 \Box\,,
\end{equation}
where $\Box$ is the d'Alambertian operator on Rindler space. The cosmological constant term in \eqref{eq:class ham} is included as a separation constant arising from the general solution to \eqref{eq: schrod eqn} and is interpreted as the negative eigenvalue of $\Box$. An equivalent evolution equation (without the self-adjoint extensions) was presented in \cite{Unruh_Wald:unimodular}, motivated by uni-modular gravity approach. This suggests that our proposal may be strongly connected to uni-modular gravity.

A theorem of Von-Neumann \cite[X.3]{reed:1975} guarantees that self-adjoint extensions of the real, symmetric operator $\hat H$ exist. Given an explicit self-adjoint representation of $\hat H$, the time evolution is guaranteed to be unitary by Stone's theorem \cite[p.264]{reed:1980}. The deficiency subspaces of $\hat H$ can be determined by computing its square integral eigenstates for the eigenvalues $\tilde \Lambda \to \pm i$. These can be found to be expressible in terms of modified Bessel functions of the second kind (see below) and have rank $(1,1)$. We therefore expect a $U(1)$ family of self-adjoint extensions, which we parametrize by the log-periodic, positive reference scale $\Lambda_\text{ref}$. To find these extensions explicitly and to construct the general solution to \eqref{eq: schrod eqn}, we compute the eigenstates of $\hat H$ (with eigenvalues $\tilde \Lambda$) in the $v\varphi$-chart. Using the separation Ansatz
\begin{equation}
  \Psi_{\Lambda}^\pm (v, \varphi) = \psi_{\Lambda, k}(v) \nu_{k}^\pm (k)\,,
\end{equation}
we find
\begin{equation}
  \nu_k^\pm(\varphi) = \frac {1}{\sqrt{2\pi\hbar}} e^{\pm \tfrac i \hbar k \varphi}\,,
\end{equation}
and
\begin{equation}\label{eq:bessel}
	v \frac{\de }{\de v}\lf( v \frac {\de}{\de v}  \psi_{\Lambda,k} \rt) + \lf( 2 \tilde \Lambda v^2 + \frac{k^2}{\hbar^2} \rt) \psi_{\Lambda,k} = 0\,.
\end{equation}
The latter equation is Bessel's differential equation for purely imaginary orders, $ik/\hbar$. 

For $\Lambda > 0$, solutions are the oscillating Bessel functions of the first, $\mathcal{J}_{ik/\hbar}$, and second kind, $Y_{ik/\hbar}$. Self-adjointness can be established by noting that near $v=0$ the Bessel functions behave as superpositions of ordinary sines and cosines of the logarithmic variable $\mu$. The phase difference,
\begin{equation}\label{eq:theta}
	\theta = \frac k {2\hbar}\log \lf( \frac {\tilde \Lambda}{\tilde \Lambda_\text{ref}(k)}\rt)
\end{equation}
between in- and out-going modes can be used to solve the appropriate boundary condition and parametrizes the $U(1)$ space of self-adjoint extensions.

The general normalized solutions are continuous in $\tilde\Lambda$ and are explicitly given by
\begin{equation}\label{eq:unbound psi}
  \psi_{\Lambda,k} 
  = \frac{ \text{\cal{Re}} \lf[  e^{-i\theta} \mathcal J_{ik/\hbar}(\sqrt{2 \tilde \Lambda} v) \rt]}{ \lf| \cosh \lf( \tfrac {\pi k}{2\hbar} + i \theta \rt)  \rt|  }\,.
\end{equation}
The $2\pi$ periodicity in $\theta$ implies a $\pi k/\hbar$ log-periodicity in $\Lambda_\text{ref}$.

For $\Lambda<0$, bound solutions can be constructed and are found to have discrete spectrum \cite{Gryb:2017a}. We will motivate excluding these solutions in the section on modelling constraints below.  

The general solution to \eqref{eq: schrod eqn} is then,
\begin{equation}\label{eq:psi final}
  \Psi(v,\varphi, t) = \sum_{\pm}\int \frac{\de k \de \Lambda}{\sqrt 2}\, e^{ -i \tilde\Lambda t/\hbar } B^\pm(\tilde \Lambda,k) \Psi^{\pm}_{\Lambda,k}\,,
\end{equation}
for the suitably normalized coefficients $B^\pm(\tilde\Lambda,k)$. Standard Wheeler--DeWitt type quantization of mini-superspace can be obtained as a special idealization of our solutions if one takes $B^\pm(\tilde\Lambda,k)$ to be a delta-function in $\tilde \Lambda$.\\

\section{Singularity Resolution} 

There are good reasons to demand that any criterion for non-singular behaviour in a quantum theory should be dynamical \cite{husain:2004,bojowald:2007,Kiefer:2007}. The most basic dynamical criterion is that a quantum theory is non-singular if the expectation value of all observable operators remains finite when evaluated on all states. It is straightforward to demonstrate that our model satisfies this criterion. Given a self-adjoint $\hat H$, \eqref{eq: schrod eqn} implies the generalised Ehrenfest theorem:
\begin{equation}\label{eq:Ehrenfest}
\frac{\partial}{\partial t} \Big{<}\hat{O}(t)\Big{>} = \frac{1}{i\hbar} \Big{<}[\hat{O}(t),\hat{H}]\Big{>}+ \Big{<}  \frac{\partial \hat{O}(t)}{\partial t}\Big{>}\,.
\end{equation} 
Provided that $\hat O$ is a self-adjoint representation of an algebra of bounded linear operators, the commutator on the RHS is also bounded and the evolution of the expectation value of all $\hat O$ will be well-behaved.\footnote{The Hamitlonian, $\hat H$, is bounded provided we restrict to $L^2$ functions on its domain; i.e., by imposing suitable falloff conditions for $B^\pm(\tilde \Lambda,k)$. It remains bounded because $\hat H$ commutes with itself.} Since our observables and Hamiltonian satisfy this criterion, the classical singularity, which results from $\pi_v \to -\infty$ when $v = 0$, is avoided by the finiteness of the corresponding quantum expectation value $\mean{\hat\pi_v} > -\infty$. In Wheeler--DeWitt-type quantizations where the Hamiltonian is treated a pure constraint, time evolution is recovered as a non-unitary operator on a reduced system. The argument for singularity resolution described above is thus only applicatble to a Schro\"odinger-type evolution equation of the form \eqref{eq: schrod eqn} and not to the Wheeler--DeWitt case.

% ==================================== Start of Introductory part ===========================================
\section{Modeling Constraints}

The choice of physically relevant particular solutions is under-constrained by observational data. Here we assume that constraints placed upon the model that are \textit{not} based upon observational data should be minimally specific: we should say as little as possible about that which we do not know. In \cite{Gryb:2017b}, this idea is articulated in terms of a principle of \emph{epistemic humility} with regards to constraining the universal wavefunction. The main conclusions of this analysis are summarised below. For a more complete justification of these parameter choices, see the details in \cite{Gryb:2017b}.

Observational data imply that the current universe is well-approximated by a semi-classical state with a definite positive $\Lambda$ with no evidence of bound negative $\Lambda$ states. This justifies our use of only $\Lambda >0$ eigenstates. We can characterise the semi-classical regime in a minimally specific way by the vanishing of higher order generalized moments of the wavefunction \cite{brizuela:2014}. This is equivalent to requiring that the non-Gaussianties of the wavefunction are very small in a particular basis. The minimally specific choice of basis is that which is maximally stable.\footnote{Ultimately, what is needed is a super-selection principle for the stable basis to be singled out by decoherence with suitable `environmental' degrees of freedom \cite{kiefer:1995}.} The large-$v$ asymptotic Killing vectors of the classical configuration space allow us to select such a stable basis given in terms of the eigenstates of $\hat\pi_\varphi$ and $\hat\pi_v$. Since in this asymptotic limit, $H = \frac 1 {2\hbar^2} \pi_v^2$, we take the semi-classical state to be expressed in terms of Gaussians of $k$ and $\omega = \sqrt{2 \tilde\Lambda} \hbar$ (the approximate eigenvalues of $\hat \pi_v$ in the large-$v$ limit). Crucially, the wavefunction will \emph{not} remain Gaussian in the basis defined by the operator $\hat v$, which will become highly non-Gaussian near the bounce. Gaussianity in the $\omega$-basis can therefore be understood as keeping the $v$-basis as semi-classical as possible throughout the evolution.

Requiring $\Lambda$ and $\pi_\varphi$ to be well-resolved implies that the absolute value of the means of the scalar densities $B^\pm(k,\tilde\Lambda)\,\de \Lambda = \frac \omega \hbar B^\pm(k,\omega)\, \de \omega$ must be much larger than the variances;  otherwise the quantum mechanical uncertainty, given by $\sigma_\omega$ and $\sigma_k$ respectively, would make them indistinguishable from zero. This leads to:
\begin{multline} \label{eq:k and E wavefunction}
  \frac \omega \hbar B^\pm(k,\omega) = \lf(  \frac{\hbar^2}{2 \pi \sigma_\omega \sigma_k}\rt)^{1/2} \exp \lf\{- \frac { (\omega- \omega_0)^2}{4 \sigma_\omega^2}\rt. \\ \lf. - \frac i \hbar (\omega- \omega_0) v_{0}  - \frac{(k- k_0^\pm)^2}{ 4 (\sigma_k^\pm)^2} - \tfrac i \hbar (k-k_0^\pm) \varphi_{\infty}^\pm \rt\} \,,
\end{multline}
where $\omega_0 \gg \sigma_\omega > 0$ and $|k_0^\pm| \gg \sigma_k^\pm > 0$.

Further minimally specific choices consistent with observation are: i) to select $t=0$ as the time of minimal dispersion by appeal to time-translational invariance; and ii) to assume a semi-classical regime for $t\rightarrow \pm \infty$. These modeling constraints encode the core features of a quantum bounce into our solutions. While states may have wildly varying behaviour before the bounce, current observational limitations are too significant to give any indications of this pre-bounce physics. One must therefore make assumptions for these early states. The minimally specific choice is to impose the maximum amount of time-reflection symmetry around the bounce. This is achieved by: iii) setting the phase shift between in- and out-going $\hat\pi_\varphi$-eigenstates to zero by setting $B^+ = B^-$ using a single mean, $k_0$, and variance; $\sigma_k$, and offset, $\varphi_\infty$; iv) requiring the bounce time to occur at $t=0$ by setting $v_0 = 0$; and v) fixing the self-adjoint extensions to minimize the phase-shift between in- and out-going $\hat H$-eigenstates (the specific choice that accomplishes this is specified below).

%and the density $E$ transforms such that $E(\Lambda) \to \frac \omega \hbar E(\omega)$. 

We can use the global `boost' isometry of Rindler space to restrict to $\varphi_\infty = 0$ without loss of generality. The parameter pairs $(k_0, \sigma_k)$ and $(\omega_0, \sigma_\omega)$ can only be independently defined via reference to an external scale. We can avoid having to specify such a scale by noticing that the Gaussians of \eqref{eq:k and E wavefunction} depend only on the ratios $k_0/\sigma_k$ and $\omega_0/\sigma_\omega$, which independent parameters of the model. 

Fixing the self-adjoint extensions by specifying $\theta$ requires the introduction of an external reference scale via its definition \eqref{eq:theta}. This reference scale can be thought of as giving meaning to the units of $\Lambda$ which are needed to make sense of its influence on the boundary. Inspection of \eqref{eq:theta} reveals that the freedom in choosing $\Lambda_\text{ref}$ can be absorbed into a choice of $k_0/\hbar$, which we choose as the third free parameter of the model. We will discuss the physical interpretation of this scale in relation to Planck-scale effects shortly. For our present purpose, we choose: 
\begin{equation}\label{eq:Lambda ref choice}
  \Lambda_\text{ref} = \frac{V_0^2}{\kappa^2 \hbar^2} \frac {\omega_0^2}{2\hbar^2}\,,
\end{equation}
which selects the natural classical units for $\Lambda$ and is minimally specific at the quantum level because it does not involve introducing any new parameters.

\section{Rayleigh limit} The Rayleigh limit is that in which the cosmological constant is well-resolved semi-classically. We can restrict our solutions to this limit by choosing the parameters of our model to satisfy the relation $\omega_0/\sigma_\omega \gg 1$. In the Rayleigh limit, Planck-scale effects will be found to be negligible in a manner analogous to the negligibility of molecular effects in Rayleigh scattering. This occurs because, when $v\omega/\hbar \gg 1$, the Bessel functions in \eqref{eq:unbound psi} take the form of cosine functions according to
\begin{equation}
	\mathcal{J}_{ik}\lf( \tfrac {\omega v}\hbar \rt) \sim \cos \lf( \tfrac {\omega v}\hbar - \tfrac \Delta 2 \rt)\,.
\end{equation}
The variables $v$ and $\omega$ thus behave as conjugate coordinates in this limit so that, for a Gaussian state, the uncertainty principle is saturated: $\sigma_v \sim \hbar/\sigma_\omega$. During the bounce, where $v$ is smallest, $v \sim \sigma_v$ and the Rayleigh limit immediately implies
\begin{equation}
  \frac{v\omega}\hbar \sim \frac{\omega_0}{\sigma_\omega} \gg 1\,.
\end{equation}
The bounce therefore occurs in a regime where the asymptotic expansion of the Bessel functions is approximately valid and the system is reasonably described by two nearly Gaussian envelopes associated with the in- and out-going $v$-space waves contained in the cosine function. During the bounce overlap between these envelopes produces interference `beats' in $v$-space with a frequency set by the size of $\omega_0/\hbar$. This implies that the number of beats in a single envelope scales like
\begin{equation}\label{eq:envelope to beats}
	\frac{\text{envelope size}}{\text{beat size}} \sim \omega_0/\sigma_\omega \gg 1\,.
\end{equation}
These features can be understood analytically in the limit $\frac{\omega_0/ \sigma_\omega}{|k_0|/ \sigma_k} \gg 1$, where the classical system resembles a de~Sitter geometry \cite{Gryb:2017b}.

\section{Bouncing Cosmology} %300 words clearly

Given the log-periodicity of $\Lambda_\text{ref}$, the limit
\begin{equation}\label{eq:universality limit}
  e^{|k_0|/\hbar} \gg e^{\omega_0/\sigma_\omega}
\end{equation}
implies that for any choice of $\Lambda_\text{ref}$ there is an equivalent one imperceptibly close to $\Lambda_0$. The behaviour of the self-adjoint extensions is thus found to be universal in this limit. Using \eqref{eq:universality limit}, the normalization of the unbound eigenstates, \eqref{eq:unbound psi}, simplifies to $\text{sech}\lf(  \frac{\pi k}{2\hbar} \rt)$, which is $\omega$-independent. The integration of \eqref{eq:psi final} over $\omega$ for the Gaussian $B(\omega,k)$ can then be well-approximated in terms of confluent hypergeometric functions \cite{Gryb:2017b}. The remaining integral reduces to a Fourier transform in $k$, which can be evaluated numerically.

To analyze the resulting solutions, we consider the effect of the three independent parameters $k_0/\hbar$, $\omega_0/\sigma_k$, and $k_0/\sigma_k$ separately. The choice of self-adjoint extension \eqref{eq:Lambda ref choice} minimizes the phase difference between in- and out-going modes due to non-zero $k_0/\hbar$. We therefore expect this choice to lead to a negligible correction to the beat frequency predicted by the considerations of the previous section in the Rayleigh limit. Numerical evidence for this can be seen by explicit comparison of the Born amplitudes of the wavefunction in the $v\varphi$-basis for modest parameter values (e.g., $\omega_0/\sigma_\omega = 10$, $k_0/\sigma_k = 10$, $\hbar = 1,2$).

\begin{figure}
  \subfloat[$v |\Psi|^2$ for $\omega_0/\sigma_\omega = 10$, $s =1$]{
    \includegraphics[width=0.85\linewidth]{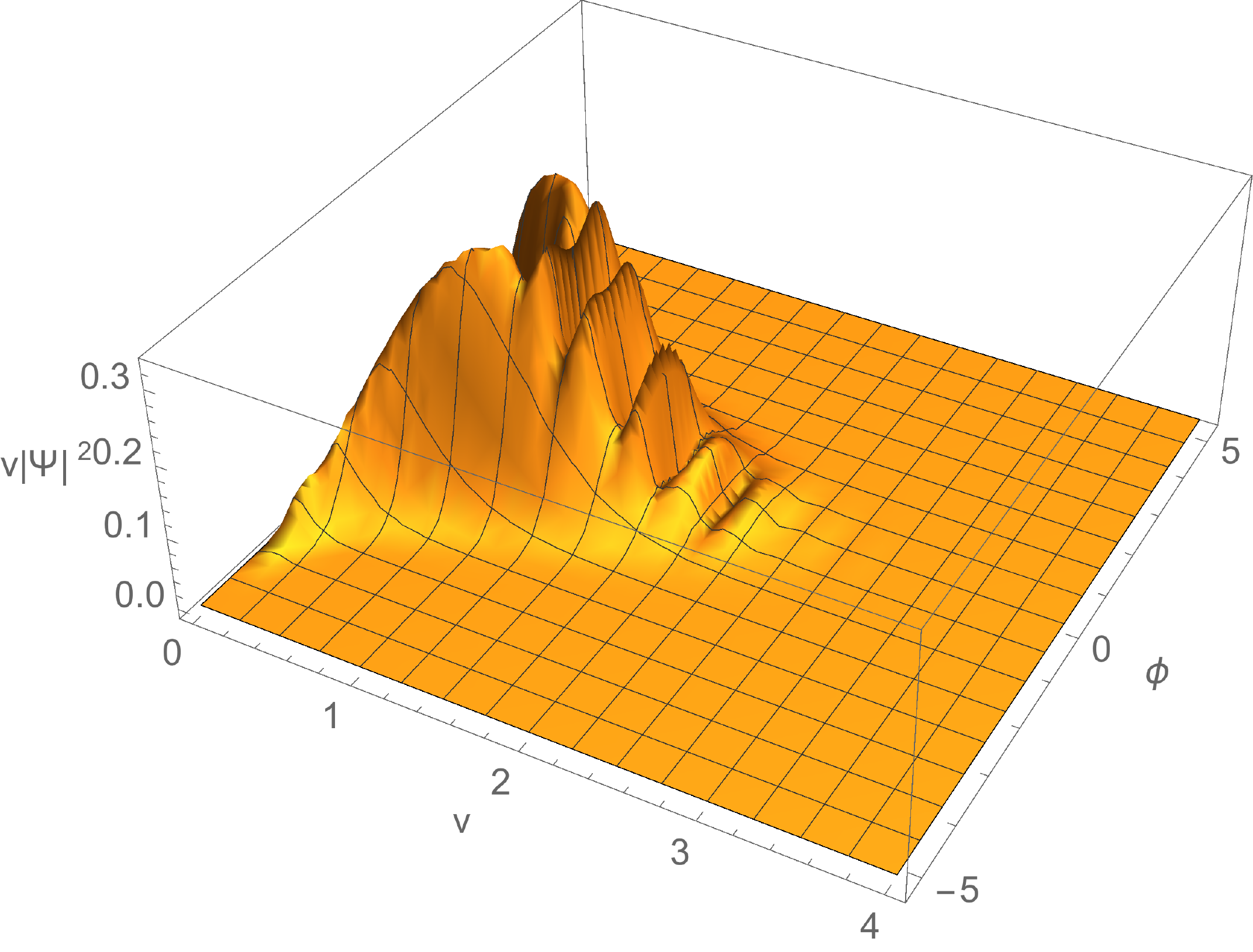}
    \label{fig:bounce_omega_10}
  }\\
  \subfloat[$v |\Psi|^2$ for $\omega_0/\sigma_\omega = 15$, $s=1$]{
    \includegraphics[width=0.85\linewidth]{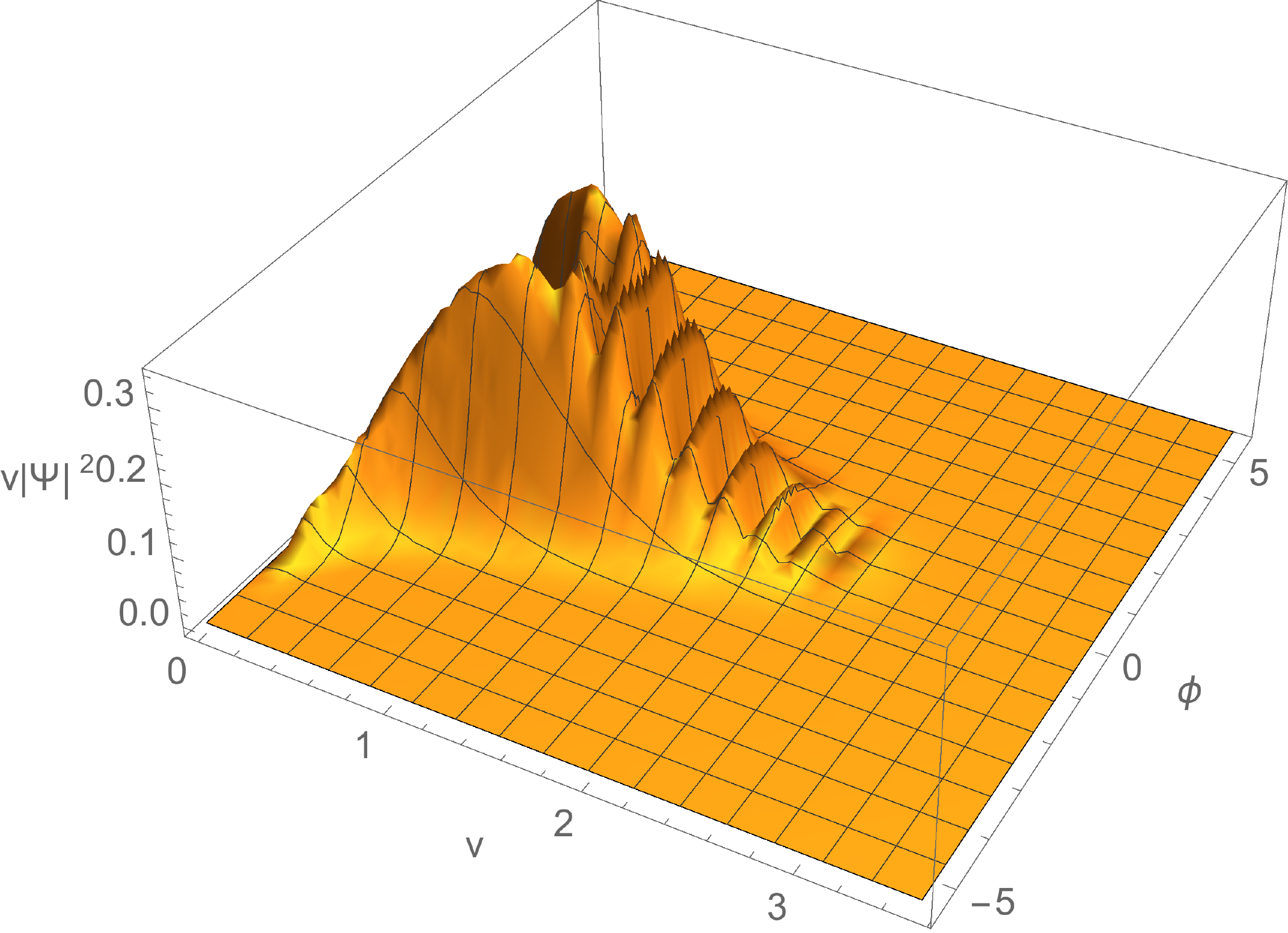}
    \label{fig:bounce_omega_15}
  }
  \caption{\label{fig:bounce_comp} Comparison of Born amplitude at bounce time for different choices of $\omega_0$ and $s$ (for $\sigma_\omega = \sigma_k= \hbar = 1$).}
\end{figure}

The parameter $\omega_0/\sigma_\omega$ is expected to control the number of beats in an envelope according to \eqref{eq:envelope to beats}. To verify this relation, we can plot (see FIG~\ref{fig:bounce_comp}) the Born amplitude of the wavefunction in the $v\varphi$-basis at $t=0$, where the overlap is maximum. Comparison of the beat frequency for different values of $\omega_0/\sigma_\omega$ is in excellent agreement with \eqref{eq:envelope to beats}.

\begin{figure}
  \includegraphics[width=\linewidth]{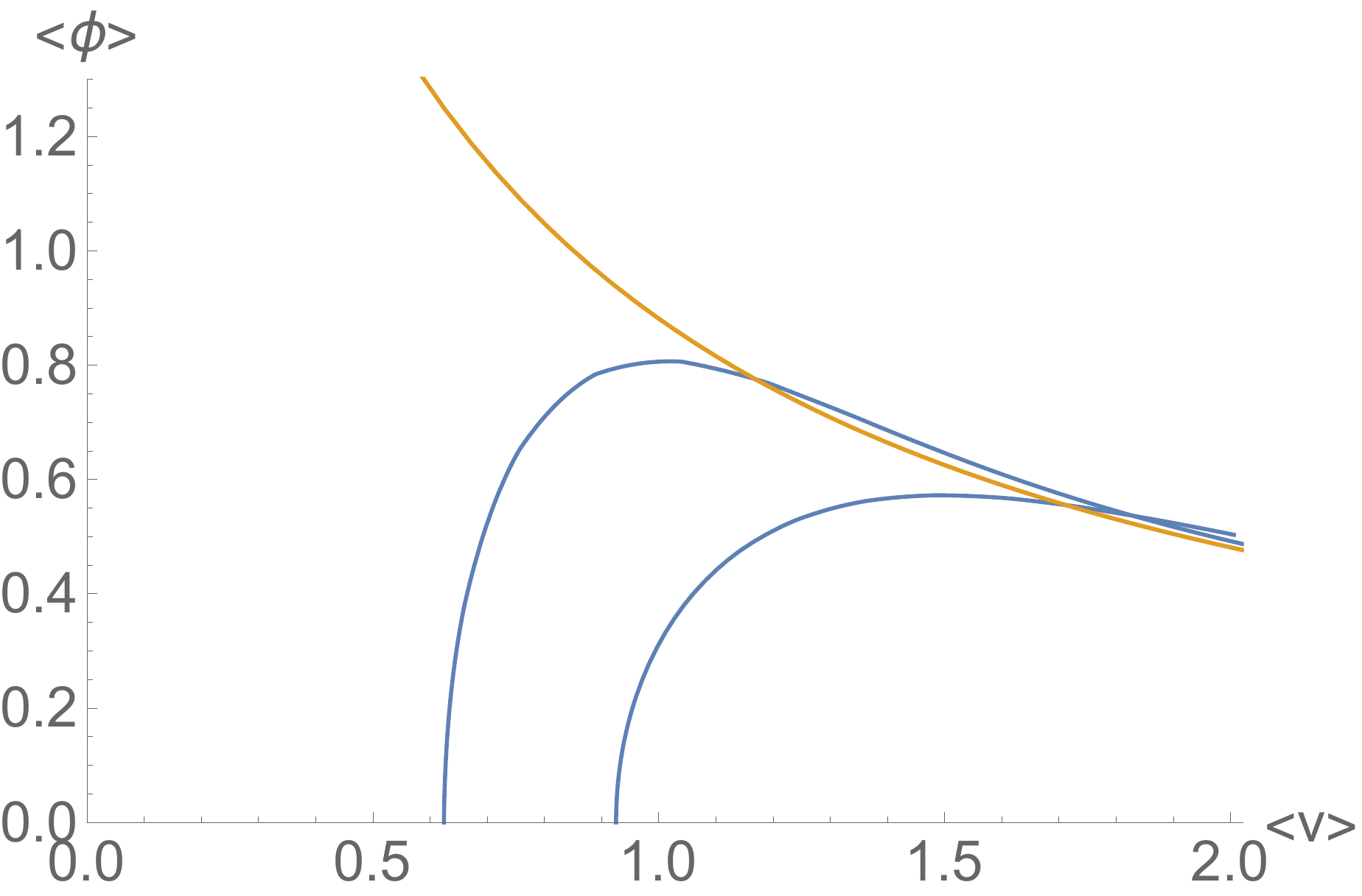}
  \caption{\label{fig:mean v mean phi} $\mean{\hat v}/s$ versus $\mean{\hat\varphi}$ for different $s$ and $\omega_0 = 10$. Top blue: $s=1$; bottom blue: $s=2$; yellow: classical. (Symmetric upon $\varphi \to -\varphi$.)}
\end{figure}

The parameter $k_0/\sigma_k$ controls how tightly the individual envelopes stay peaked on the classical solutions. This can be studied by varying the parameter $s = k_0/\omega_0$ for fixed $\omega_0/\sigma_\omega$ and $k_0/\hbar$ and parametrically plotting $\mean{\hat v}/s$ and $\mean{\hat \varphi}$. The advantage of this choice of parameterization of the quantum solutions in terms of $s$ is that the classical equations of motion can be written parametrically as
\begin{equation}
  \frac v s = \lf| \text{cosech}\lf( \varphi - \varphi_\infty \rt)  \rt|\,.
\end{equation}
The quantum curve for different choices of $s$ can thus be compared with the same universal classical curve. FIG~\ref{fig:mean v mean phi} illustrates the relevant features. The expectation values begin to diverge from their classical values in the region $v \sim 1/\sigma_\omega$ as expected. The expectation value of $\hat\varphi$ reaches a maximum value, which increases as $s$ increases. The expectation value of $\hat v$ reaches a minimum at $t = 0$ as expected.

\section{Prospectus}

Let us briefly consider potential connection between the model and  inflationary cosmology. In particular, following \cite{PhysRevD.42.3936,Liddle:1994dx}, it is possible to define \emph{effective} slow-roll parameters using the implicit dependence of the expectation value of the Hubble parameter as a function of $\mean{\hat\varphi}$. This effective Hubble parameter is proportional to $\mean{\hat\pi_v}$ and therefore the relevant slow-roll parameters are proportional to the first and second derivatives of $\mean{\hat\pi_v}$. Because $\mean{\hat\pi_v}$ is bounded, the fact that it is zero at the bounce and decreasing to a positive constant at late times implies that it must have a maximum at some point in the epoch that takes place after the bounce and before the semi-classical regime. This maximum should be similar to the maximum seen in $\mean{\hat\varphi}$ in FIG.~\ref{fig:mean v mean phi}. Near this maximum, the condition for the first slow-roll parameter is satisfied. Because it is possible in our model to arbitrarily widen the width of the wavefunction in $v$-space near the bounce by taking increasingly narrow distributions of the cosmological constant, there may be enough freedom in the parameter space of our model to extend the second rate of change of $\mean{\hat\pi_v}$ near this maximum so that the condition for the second slow-roll parameter is also satisfied. This would be similar to how the second rate of change of $\mean{\hat\varphi}$ can be slowed near its maximum by increasing $s$ as can be seen in FIG.~\ref{fig:mean v mean phi}. The Rayleigh limit also suggests that this potential inflationary regime could be found to take place far below the Planck energy. This raises the exciting possibility, to be pursued in future work, of describing inflationary scenarios using the model.

A further extension of our work is a unitary quantization of anisotropic Bianchi models \cite{wainwright_ellis_1997}. While the extension to Bianchi~I is straightforward, Bianchi~IX models will lead to modifications of the Bessel equations. This notwithstanding, one may expect that many of the qualitative features of the present model will carry forward to solutions of the Bianchi~IX model that persist to the late-time attractors. The Bianchi~IX model may be particularly valuable for studying general singularity resolution in quantized GR in light of the BKL conjecture \cite{Belinsky:1970ew}. Such a framework may be useful for studying singularity resolution of time-like singularities via, for example, black-to-white hole transitions.

% -------------------------------------- Acknowledgments ---------------------------------------
\section{Acknowledgments} % (fold)
\label{sec:acknowledgments}

   We are appreciative to audiences in Bristol, Berlin, Geneva, Harvard, Hannover, Nottingham and the Perimeter Institute for comments. We also thank Henrique Gomes, David Sloan, and Martin Bojowald for helpful comments. We are very grateful for the support from the Institute for Advanced Studies and the School of Arts at the University of Bristol and to the Arts and Humanities Research Council (Grant Ref. AH/P004415/1). S.G. would like to acknowledge support from the Netherlands Organisation for Scientific Research (NWO) (Project No. 620.01.784) and Radboud University. K.T. would like to thank the Alexander von Humboldt Foundation and the Munich Center for Mathematical Philosophy (Ludwig-Maximilians-Universit\"{a}t M\"{u}nchen) for supporting the early stages of work on this project. 

% ---------------------------------------- Appendices (optional) -------------------------------

%\appendix     % For many appendix sections
%\appendix*    % For a single appendix section.

% subsection unbound_eigenstates (end)

% --------------------------------------- Bibliography ------------------------------------------

%\bibliographystyle{utphys}
%\bibliography{mach,Masterbib,Masterbib2}

\providecommand{\href}[2]{#2}\begingroup\raggedright\endgroup

\end{document}